\documentclass[prb,twocolumn,showpacs,preprintnumbers,amsmath,amssymb]{revtex4}
\usepackage{graphicx}
\usepackage{dcolumn}
\usepackage{bm}

\begin{document}

\title{Non-zero macroscopic  magnetization in half-metallic
antiferromagnets \\ at finite temperatures}

\author{Ersoy  \c{S}a\c{s}{\i}o\u{g}lu}\email{e.sasioglu@fz-juelich.de}
\affiliation{Institut f\"{u}r Festk\"{o}rperforschung,
Forschungszentrum J\"{u}lich, D-52425 J\"{u}lich, Germany \\ and
Department of Physics, Fatih University, B\"{u}y\"{u}k\c{c}ekmece,
TR-34500, \.{I}stanbul, Turkey}

\date{\today}

\begin{abstract}

Combining density functional theory calculations with many-body
Green function technique, we reveal that the macroscopic
magnetization in half-metallic antiferromagnets does not vanish at
finite temperature as for the $\textrm{T}=0$ limit. This anomalous
behavior stems from the inequivalent magnetic sublattices which
lead to different intra-sublattice exchange interactions. As a
consequence, the spin fluctuations suppress the magnetic order of
the sublattices in a different way leading to a ferrimagnetic
state at finite temperatures. Computational results are presented
for the half-metallic antiferromagnetic CrMnZ
($\textrm{Z}=\textrm{P}$, As, Sb) semi-Heusler compounds.

\end{abstract}


\maketitle

Half-metallic antiferromagnets (HM-AFMs) are considered to be the
most promising class of materials for spintronics
applications.\cite{Pickett_1,Katsnelson} A HM-AFM material is not
antiferromagnetic in the usual sense of the term; it is a special
case of a ferrimagnet with compensated sublattice magnetization.
The existence of the gap in one of the spin channels (either up or
down) leads to the complete cancellation of the magnetic moments
at zero temperature with a 100\% spin polarization of the charge
carriers at the Fermi level. In conventional AFMs the macroscopic
spin polarization is zero due to the spin rotational symmetry with
the exception of the compounds with broken inversion symmetry like
$\alpha-$Fe$_{2}$O$_{3}$ in which spin-orbit gives rise to weak
ferromagnetism (0.002 $\mu_B$) due to the canting of the magnetic
moments.\cite{weak_FM1,weak_FM2} The HM-AFM materials provide
several advantages in device applications with respect to the
half-metallic ferromagnets (HM-FMs). For example, they would be
perfectly stable spin-polarized electrodes in a junction device.
These materials do not give rise to stray fields, and thus no
magnetic domain walls are formed. Besides applications in
spintronics, HM-AFMs provide a possibility of "single spin
superconductivity" due to the spin triplet ($S=1$) pairing in
metallic channel.\cite{Pickett_2}

The possible existence of HM-AFM was pointed out by van Leuken and
de Groot in 1995.\cite{Leuken}  Based on first-principles
calculations authors proposed CrMnSb and
V$_{7}$MnFe$_{8}$Sb$_{7}$In  as candidates for HM-AFMs. Later,
Pickett suggested that also the cubic double perovskites
La$_2$VCuO$_6$, La$_2$MnVO$_6$, and La$_2$MnCoO$_6$ are
HM-AFMs.\cite{Pickett_3}  Since then substantial effort has been
devoted to find materials with  HM antiferromagnetic
characteristics. Predicted promising systems include double
perovskites\cite{DP_1,DP_2}, thiospinels\cite{TS}, tetrahedrally
coordinated transition-metal-based chalcopyrites\cite{Nakao_1},
full-Heusler alloys\cite{Felser,Galanakis_1} and monolayer
superlattices.\cite{Nakao_2}  Not only ordered but also several
disordered systems have been shown to be
HM-AFMs.\cite{Galanakis_2,RS-TMO,Cuprates,DMS_1,DMS_2,DMS_3}  In
spite of substantial theoretical efforts for designing materials
with HM-AFM characteristics and the study of their ground state
electronic and magnetic properties, only few works exist
addressing the exchange interactions  and magnetic phase
transition temperatures in these systems.\cite{DMS_2,DMS_3}

In this Communication we investigate the temperature dependence of
the magnetization in HM-AFMs employing the many-body Green
function technique\cite{MBGFT} within Tyablikov decoupling
scheme.\cite{Tyablikov} For computational purposes we consider
CrMnZ ($\textrm{Z}=\textrm{P}$, As, Sb) semi-Heusler compounds
which are the simplest systems among the predicted HM-AFMs with
two magnetic atoms per unit cell and which are compatible with the
existing semiconductors technology.  However, present findings are
valid for more complicated systems like double
perovskites\cite{DP_1,DP_2} or diluted antiferromagnetic
semiconductors.\cite{DMS_2,DMS_3} We show that the macroscopic
magnetization in these materials does not vanish at finite
temperature in contrast to the zero temperature limit and
conventional AFMs. This peculiar behavior originates from the
inequivalent magnetic sublattices in HM-AFMs which lead to
different intra-sublattice exchange interactions and, as a
consequence, spin fluctuations suppress the magnetic order of the
sublattices in a different way. Thus, at finite temperature
sublattice magnetizations do not compensate each other and all
three compounds show ferrimagnetic behavior which seems to be
contradictory to our knowledge on finite temperature properties of
the magnetic materials. However, this seemingly counterintuitive
results can be explained by an analysis of the electronic
structure and exchange interactions in these systems.

Ground state calculations are carried out using the augmented
spherical wave method within the generalized gradient
approximation to the exchange correlation potential. Details of
the computational scheme can be found in
Ref.~\onlinecite{method_1}. To provide the basis for further
considerations we start with a brief discussion of the electronic
structure of the CrMnZ compounds. Like several Heusler alloys
these systems have theoretical equilibrium lattice constants (see
Table~\ref{table1}) close to the ones of  the zincblende
semiconductors (GaP, GaAs). Thus, calculational results will be
presented for the latter case since these semiconductors might be
considered as possible substrates to grow these
materials.\cite{lattice}  All compounds under study have 18
valance electrons per unit cell and calculations show that the
total spin moments given in Table~\ref{table1} are exactly zero in
agreement with the Slater-Pauling behavior for ideal
half-metals.\cite{Galanakis_Review} Simultaneously, the spin
magnetic moments of Cr and Mn atoms are antiparallel and these
compounds are ferrimagnets with compensated sublattice
magnetizations. The origin of the HM gap in Heusler alloys has
been well understood  and the reader is referred to
Ref.~\onlinecite{Galanakis_Review} since the same discussion is
valid for the present systems.

\begin{table}
\begin{center}
\caption{Lattice parameters and spin magnetic moments (in $\mu_B$)
for CrMnZ ($\textrm{Z}=\textrm{P}$, As, Sb). The calculated
equilibrium lattice constants are 5.44 $\AA$, 5.71 $\AA$ and 6.08
$\AA$ for $\textrm{Z}=\textrm{P}$, As and Sb, respectively, close
to the experimental lattice parameters of GaP, GaAs and InP.}
\begin{ruledtabular}
\begin{tabular}{llcccc}
 Compound  & a($\AA$) & m$_{\textrm{Cr}}$  &  m$_{\textrm{Mn}}$  &    m$_{\textrm{Z}}$  & m$_{\textrm{Cell}}$ \\ \hline

     CrMnP    & $5.45_{\textrm{[GaP]}}$   & 1.80 & -1.83 &    0.03 & 0.00   \\
     CrMnAs   & $5.65_{\textrm{[GaAs]}}$  & 2.52 & -2.54 &    0.02 & 0.00   \\
     CrMnSb   & $5.87_{\textrm{[InP]}}$   & 2.74 & -2.76 &    0.02 & 0.00   \\

\end{tabular}
\end{ruledtabular}
\label{table1}
\end{center}
\end{table}

To study interatomic exchange interactions  we map the complex
itinerant electron problem onto a classical Heisenberg Hamiltonian
$H=-\sum_{\mu,\nu}\sum_{\textbf{R},\textbf{R}'}
J^{\mu\nu}_{\textbf{R}\textbf{R}'}s^{\mu}_{\textbf{R}}s^{\nu}_{\textbf{R}'}$
where $\mu\textbf{R}\neq \nu\textbf{R}'$ and the indices $\mu$ and
$\nu$ represent different sublattices. \textbf{R} and
$\textbf{R}'$ are the lattice vectors specifying the atoms within
the sublattices, and $s^{\mu}_{\textbf{R}}$ is the unit vector
pointing in the direction of the magnetic moment at site
($\mu$,\textbf{R}). Heisenberg exchange parameters
$J^{\mu\nu}_{\textbf{R}\textbf{R}'}$ are calculated employing the
frozen-magnon technique as described in
Ref.~\onlinecite{method_1}. Extensive investigations on the
multi-sublattice Heusler alloys have shown that there are several
exchange interactions which coexist and which are mixed
together.\cite{method_1,Rusz,Kudrnovsky,method_2} To simplify the
discussion let us write the total magnetic exchange field acting
on the sublattice $\mu$ as $J_{\textmd{total}}^{\mu} \sim
J_{\textmd{direct}}^{(\mu\nu)} + J_{\textmd{indirect}}^{(\mu\nu)}
+ J_{\textmd{indirect}}^{(\mu\mu)}$ where the first two terms
represent the direct and indirect exchange coupling between
different sublattices. The former (direct coupling) provides the
leading contribution to the total exchange coupling and determines
the character of the magnetic state while the latter contributes
little and is not so important. On the other hand, the last term,
i.e., the intra-sublattice coupling is of particular importance
because it is responsible for the appearance of a net macroscopic
magnetization at finite temperature in HM-AFMs. It should be noted
here that in reality the situation is not so simple, the exchange
field acting on the sublattices should be determined from the
solution of a matrix equation.

\begin{figure}[t]
\begin{center}
\includegraphics[scale=0.5]{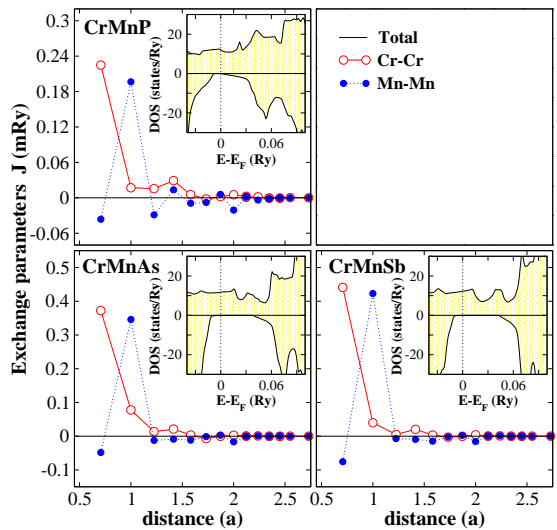}
\end{center}
\vspace*{-0.6cm} \caption{(Color online) Intra-sublattice Cr-Cr
and Mn-Mn exchange interactions in CrMnZ ($\textrm{Z}=\textrm{P}$,
As, Sb) as a function of distance. In the insets we show
spin-resolved total density of states around Fermi level.}
\label{fig1}
\end{figure}

\begin{figure}[t]
\begin{center}
\includegraphics[scale=0.5]{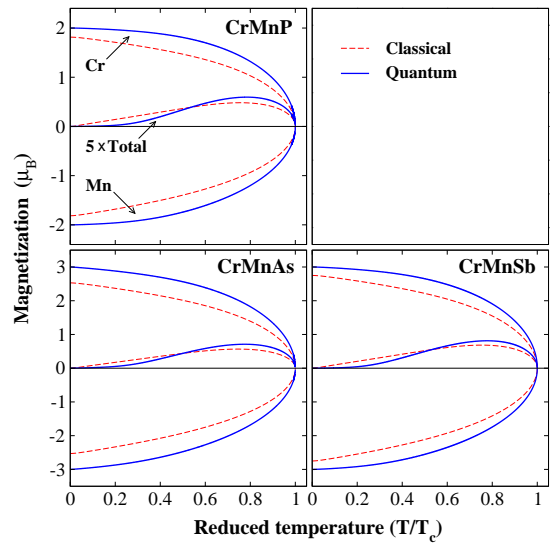}
\end{center}
\vspace*{-0.6cm} \caption{(Color online) The calculated
temperature dependence of the sublattice and  total magnetization
for CrMnZ ($\textrm{Z}=\textrm{P}$, As, Sb). The calculations are
performed for both classical and quantum Hamiltonians. The
temperature is given in reduced form and the total magnetization
is scaled up by a factor of 5. } \label{fig2}
\end{figure}

Let us start with the discussion of the last term. Because of the
large distance between Cr-Cr (Mn-Mn) atoms this coupling is
indirect mediated by the conduction electrons. In Fig.~\ref{fig1}
we present the calculated Cr-Cr and Mn-Mn Heisenberg exchange
parameters as a function of the distance. As seen, due to
inequivalent magnetic sublattices the pattern of Cr-Cr and Mn-Mn
exchange parameters show very different behavior. Although the
former has ferromagnetic character, the latter shows oscillatory
behavior so that the ferromagnetic and antiferromagnetic
contributions partly compensate each other giving rise to a small
contribution into the total exchange field acting on the Mn
sublattice. In Table~\ref{table2} we present the intra-sublattice
($J_{0}^{\mu}=\sum_{\textbf{R}\neq 0}J_{0\textbf{R}}^{\mu\mu}$) as
well as the inter-sublattice
($J_{0}^{\mu\nu}=\sum_{\textbf{R}}J_{0\textbf{R}}^{\mu\nu}$)
on-site exchange coupling parameters. The on-site Cr-Cr and Mn-Mn
exchange couplings are rather different and this difference will
be reflected as a net macroscopic magnetization at finite
temperatures (see Fig.~\ref{fig2}). It is worth to note that in
conventional AFMs this coupling is the same for both sublattices.
The increase of the strength of the exchange interactions and
correspondingly of the critical temperatures in the P-As-Sb
sequence can be explained by the increase of the magnetic moments
(see Table~\ref{table1}). Moreover, due to the presence of the HM
gap the exchange interactions quickly decay.\cite{Rusz,Kudrnovsky}

In contrast to the intra-sublattice exchange interactions, the
inter-sublattice (Cr-Mn) ones behave very differently. Due to the
smaller Cr-Mn distance a very strong antiferromagnetic
nearest-neighbor direct interaction takes place between these
atoms which is about one order of magnitude larger than the
nearest-neighbor Cr-Cr coupling and is responsible for the
formation of the ferrimagnetic state. The interactions between
further nearest neighbors are very small. The on-site Cr-Mn
exchange couplings presented in Table~\ref{table2} are about 4 and
20 times larger than the Cr-Cr and Mn-Mn ones, respectively. Note
that the ferromagnetic intra-sublattice exchange interactions
further stabilize the ferrimagnetic order. The antiferromagnetic
coupling of the Cr and Mn atoms can be qualitatively explained on
the basis of the following facts: First, half-filled shells tend
to yield a strong trend towards antiferromagnetism and second,
exchange coupling in 3\textit{d} transition metals obeys the
semi-phenomenological Bethe-Slater-N\'{e}el curve which predicts
antiferromagnetism in the case of small interatomic
distances.\cite{Bethe_Slater} Indeed, the Cr-Mn distance
($d_{[\textmd{Cr-Mn}]}=2.36-2.54 \AA$) in the present systems is
comparable with the Cr-Cr distance ($d_{[\textmd{Cr-Cr}]}=2.52
\AA$) in the antiferromagnetic bcc Cr and both magnetic atoms
posses half-filled 3\textit{d} shells, thus, antiferromagnetic
Cr-Mn coupling is expected.

\begin{table}
\begin{center}
\caption{On-site exchange parameters (in mRy)  and estimated
critical temperatures (in K) within RPA  for quantum
(T$_{\textrm{C}}^{\textrm{[Q]}}$) and classical
(T$_{\textrm{C}}^{\textrm{[C]}}$) spins for CrMnZ
($\textrm{Z}=\textrm{P}$, As, Sb).}
\begin{ruledtabular}
\begin{tabular}{llcccc}
 Compound  &  $\textrm{J}_{0}^{\textrm{[Cr-Cr]}}$ &
 $\textrm{J}_{0}^{\textrm{[Mn-Mn]}}$ & $\textrm{J}_{0}^{\textrm{[Cr-Mn]}}$ &
   T$_{\textrm{C}}^{\textrm{[Q]}}$(K) &  T$_{\textrm{C}}^{\textrm{[C]}}$(K)
\\ \hline

     CrMnP    & 3.80  & 0.03 & -12.99& 2530  & 1264  \\
     CrMnAs   & 5.55  & 0.84 & -19.50& 2610  & 1566  \\
     CrMnSb   & 5.96  & 1.06 & -20.16& 2986  & 1792  \\

\end{tabular}
\end{ruledtabular}
\label{table2}
\end{center}
\end{table}

With calculated exchange parameters in hand, now we can study
temperature dependence of the magnetization employing  the methods
of statistical mechanics to the Heisenberg Hamiltonian. We use the
many-body Green function technique\cite{MBGFT} within Tyablikov
decoupling scheme\cite{Tyablikov} [also known as the random-phase
approximation (RPA)] as described in Ref.~\onlinecite{method_2}.
Note that RPA takes into account only transverse spin fluctuations
and the spin-flip Stoner excitations (longitudinal spin
fluctuations) are neglected. However, available experimental data
on Heusler alloys have shown that these latter excitations are
well separated in energy from the former one (spin waves) due to
large exchange splitting $\Delta$ ($\Delta \sim 2-3$
eV).\cite{Webster} In addition to this the presence of HM gap
prevents spin-flip transitions. Thus, Stoner excitations do not
not play an important role in thermodynamics of the present
systems and the RPA method is well grounded. We consider both
classical-spin and quantum-spin cases. In the classical-spin
calculations the obtained  values of the magnetic moments (see
Table~\ref{table1}) are used while for the quantum mechanical case
we assign integer values to the atomic moments: $2\mu_B$ ($S=1$)
(for Cr and Mn) in CrMnP and $3\mu_B$ ($S=3/2$) in CrMnAs and
CrMnSb. In quantum-spin case the thermal average of the sublattice
magnetization is given by the Callen's expression\cite{Callen}
\begin{equation}
\langle
\hat{s}_{\mu}^{z}\rangle=\frac{(S_{\mu}-\Phi_{\mu})(1+\Phi_{\mu})^{2S_{\mu}+1}
+(S_{\mu}+1+\Phi_{\mu})\Phi_{\mu}^{2S_{\mu}+1}}
{(1+\Phi_{\mu})^{2S_{\mu}+1}-(\Phi_{\mu})^{2S_{\mu}+1}} \nonumber
\end{equation}
where $\Phi_{\mu}$ is an auxiliary function. As the $\Phi_{\mu}$
depends on the  $\langle \hat{s}_{\mu}^{z}\rangle$ as well as
$\langle \hat{s}_{\nu}^{z}\rangle$  to be determined, the above
equation forms a self-consistency problem to be solved by
iteration. Note that for classical spins the Callen's expression
is reduced to the Langevin function.\cite{method_2} In
Fig.~\ref{fig2} we present the calculated temperature dependence
of the sublattice and total magnetization for CrMnZ
($\textrm{Z}=\textrm{P}$, As, Sb). The temperature is given in
reduced form and the total magnetization is scaled up by a factor
of 5.  As seen, the spin fluctuations suppress the magnetic order
of the sublattices differently, i.e., the magnetization of the Mn
sublattice decreases faster than the Cr one and as a result the
total magnetization does not vanish at finite temperature in
contrast to the $\textrm{T}=0$ limit. This behavior can be traced
back to the different exchange fields acting on the sublattices
(see Table~\ref{table2}). The total magnetization shows
non-monotonous behavior i.e., first, it increases with increasing
temperature up to 0.8 T$_{\textmd{C}}$ and then decreases much
faster and finally becomes zero at the critical point. For quantum
mechanical case the calculated total magnetic moment around 0.8
T$_{\textmd{C}}$ is 0.1 - 0.16$\mu_B$ (0.006 - 0.01$\mu_B$ around
room temperature) which is about two orders of magnitude larger
than the spin-orbit coupling induced weak magnetic moment (0.002
$\mu_B$) in $\alpha-$Fe$_{2}$O$_{3}$.\cite{weak_FM1} We should
also note that in these systems not only the ideal case of zero
macroscopic magnetization but also half-metallicity is limited to
the $\textrm{T}=0$.\cite{Katsnelson,Dowben,Lezaic}

The nature of the spin (quantum or classical) plays an important
role in the temperature behavior of the magnetization curves. In
the quantum case, the magnetization drops slower than the
classical case and thus, the calculated T$_{\textmd{C}}$ values
are larger by a factor of $(S+1)/S$ entering the RPA expression
(see Ref.~\onlinecite{method_2}). Note that this factor becomes
unity for classical spins ($(S+1)/S \rightarrow 1$ for $S
\rightarrow \infty$). Another important point which is outside the
scope of the present work is that in both treatments we use the
exchange parameters estimated within the picture of classical
atomic moments. However, it is possible that the values of the
exchange parameters must be modified for the use in the
quantum-mechanical calculations. In general, the classical
calculation provides reasonable values of the critical temperature
compared with experiment while the quantum mechanical treatment
gives better form of the temperature dependence of the
magnetization.\cite{method_2} The calculated critical temperatures
within RPA  are presented in Table~\ref{table2}. We notice that
the predicted T$_{\textmd{C}}$ values of the CrMnAs and CrMnSb are
even higher than the fcc Co which possesses the highest critical
temperature (1400 K) among all known magnetic materials. This is
not surprising, because available experimental and theoretical
data have shown that the critical temperatures of HM ferromagnets
(or ferrimagnets) scales linearly with the average value of the
magnetic moment per atom in the unit cell.\cite{Kubler2007} In
this respect Co$_2$FeSi possesses the largest average magnetic
moment per atom of 1.5$\mu_B$  with an experimental
T$_{\textmd{C}}$ of 1100 K. However, in CrMnAs (CrMnSb) the
average value of the absolute magnetic moment per atom is
1.68$\mu_B$ (1.83$\mu_B$) which is larger than the corresponding
value in Co$_2$FeSi, and thus such high critical temperatures are
expected. Finally we should note that so far discussion is based
on the assumption that CrMnZ compounds possess
C1$_{\textrm{b}}$-type ordered crystal structure and  the effect
of disorder is completely ignored. However, in reality disorder
exist in various forms like defects, anti-sites or atomic swaps
which reduce not only the spin polarization at the Fermi level but
also the magnetic phase transition temperature.\cite{Abrikosov}
For example migration of the Cr atoms to the vacant sublattice is
expected to reduce T$_{\textmd{C}}$ substantially since as shown
in Ref.~\onlinecite{Galanakis_1} the L$2_{1}$-type Cr$_2$MnZ
compounds have T$_{\textmd{C}}$ values around room temperature.

In summary, we have studied the temperature dependence of the
magnetization in HM-AFM CrMnZ ($\textrm{Z}=\textrm{P}$, As, Sb)
compounds employing the many-body Green function technique. We
have shown that the macroscopic magnetization in these systems
does not vanish at finite temperature in contrast to the
$\textrm{T}=0$ limit. This anomalous behavior stems from the
inequivalent magnetic sublattices in HM-AFMs which lead to
different intra-sublattice exchange interactions and, as a result,
spin fluctuations suppress the magnetic order of the sublattices
in a different way. Thus, at finite temperatures, the sublattice
magnetizations do not compensate each other and all three
compounds show ferrimagnetic behavior. Moreover, the combination
of large HM gaps, high T$_{\textmd{C}}$ values and very small
macroscopic magnetization around room temperature makes CrMnAs and
CrMnSb promising candidates for spintronics applications.

%------------------------------------------------------------------

Fruitful discussions with L. Sandratskii, I. Galanakis and Ph.
Mavropoulos are  greatly acknowledged.

%----------------------------------------------------------------


\begin{thebibliography}{99}

\bibitem{Pickett_1}
Warren E. Pickett and Jagadeesh S. Moodera, Physics Today
\textbf{54}, 39 (2001).



\bibitem{Katsnelson}
M. I. Katsnelson, V. Yu. Irkhin, L. Chioncel, and  A. I.
Lichtenstein, and R.A. de Groot, Rev. Mod. Phys. \textbf{80}, 315
(2008).


\bibitem{weak_FM1}
L. M. Sandratskii and J. K\"{u}bler, Europhys. Lett. \textbf{33},
447 (1996).


\bibitem{weak_FM2}
V. V. Mazurenko, V. I. Anisimov, Phys. Rev. B \textbf{71}, 184434
(2005).



\bibitem{Pickett_2}
Warren E. Pickett, Phys. Rev. Lett. \textbf{77}, 3185 (1996).


\bibitem{Leuken}
H. van Leuken and R. A. de Groot, Phys. Rev. Lett. \textbf{74},
1171 (1995).


\bibitem{Pickett_3}
Warren E. Pickett, Phys. Rev. B \textbf{57}, 10613 (1998).


\bibitem{DP_1}
J. H. Park, S. K. Kwon, and B. I. Min, Phys. Rev. B \textbf{65},
174401 (2002).


\bibitem{DP_2}
Y. K. Wang and G. Y. Guo, Phys. Rev. B \textbf{73}, 064424 (2006).


\bibitem{TS}
M. S. Park, S. K. Kwon, and B. I. Min, Phys. Rev. B 64, 100403(R)
(2001).

\bibitem{Nakao_1}
M. Nakao, Phys. Rev. B \textbf{74}, 172404 (2006).


\bibitem{Felser}
S. Wurmehl, H. C. Kandpal, G. H. Fecher, and C. Felser, J. Phys.:
Condens. Matter \textbf{18}, 6171 (2006).



\bibitem{Galanakis_1}
I. Galanakis, K. \"{O}zdo\u{g}an, E. \c{S}a\c{s}{\i}o\u{g}lu, and
B. Akta\c{s}, Phys. Rev. B \textbf{75}, 172405 (2007).


\bibitem{Nakao_2}
M. Nakao, Phys. Rev. B \textbf{77}, 134414 (2008).



\bibitem{Galanakis_2}
I. Galanakis, K. \"{O}zdo\u{g}an, E. \c{S}a\c{s}{\i}o\u{g}lu, and
B. Akta\c{s}, Phys. Rev. B \textbf{75}, 092407 (2007).


\bibitem{RS-TMO}
D. K\"{o}dderitzsch, W. Hergert, Z. Szotek, and W. M. Temmerman,
Phys. Rev. B \textbf{68}, 125114 (2003).


\bibitem{Cuprates}
Yung-mau Nie and Xiao Hu, Phys. Rev. Lett. \textbf{100}, 117203
(2008).


\bibitem{DMS_1}
H. Akai and M. Ogura, Phys. Rev. Lett. \textbf{97}, 026401 (2006).


\bibitem{DMS_2}
L. Bergqvist and P. H. Dederichs,  J. Phys.: Condens. Matter
\textbf{19}, 216220 (2007).


\bibitem{DMS_3}
M. Ogura, C. Takahashi, and H. Akai, J. Phys.: Condens. Matter
\textbf{19}, 365226 (2007).



\bibitem{MBGFT}
P. Fr\"{o}brich  and P. J. Kuntz, Physics Reports, \textbf{432},
223 (2006)


\bibitem{Tyablikov}
S. V. Tyablikov, Methods of Quantum Theory of Magnetism (Plenum
Press, New York, 1967).



\bibitem{method_1}
E. \c{S}a\c{s}{\i}o\u{g}lu, L. M. Sandratskii, and P. Bruno, Phys.
Rev. B \textbf{70}, 024427 (2004).


\bibitem{lattice}
Among the considered compounds only CrMnSb is experimentally
synthesized with a crystal structure different than the C1$_{b}$
type semi-Heusler one [J. H. Wijngaard \textit{et al.}, Phys. Rev.
B \textbf{45}, 5395 (1992)]. However, state-of-the-art
experimental methods for the synthesis of materials such as the
molecular-beam epitaxy make possible the growth of these compounds
in metastable phases as thin films or multilayers where the
substrate determines the lattice parameter.




\bibitem{Galanakis_Review}
I. Galanakis and Ph. Mavropoulos, J. Phys.: Condens. Matter
\textbf{19}, 315213 (2007).


\bibitem{Rusz}
J. Rusz, L. Bergqvist, J. Kudrnovsk\'{y}, and I. Turek, Phys. Rev.
B \textbf{73}, 214412 (2006).

\bibitem{Kudrnovsky}
J. Kudrnovsk\'{y}, V. Drchal, I. Turek, and P. Weinberger, Phys.
Rev. B \textbf{78}, 054441 (2008).


\bibitem{Bethe_Slater}
Ralph Skomski, Simple Models of Magnetism, (Oxford University
Press, 2008).



\bibitem{Webster}
P. J. Webster and K. R. A. Ziebeck, in \textit{Alloys and
Compounds of d-Elements with Main Group Elements}, Part 2, edited
by H. R. J. Wijn, Landolt-B\"{o}rnstein, New Series, Group III,
Vol. 19, part. C (Springer, Berlin, 1988).




\bibitem{Callen}
H. B. Callen, Phys. Rev. \textbf{130}, 890 (1963).


\bibitem{method_2}
E. \c{S}a\c{s}{\i}o\u{g}lu, L. M. Sandratskii, P. Bruno and I.
Galanakis, Phys. Rev. B \textbf{72}, 184415 (2005).


\bibitem{Dowben}
P. A. Dowben and R. Skomski, J. Appl. Phys. \textbf{93}, 7948
(2003).

\bibitem{Lezaic}
M. Le\v{z}ai\'{c}, Ph. Mavropoulos, J. Enkovaara, G. Bihlmayer,
and S. Blügel Phys. Rev. Lett. \textbf{97}, 026404 (2006).


\bibitem{Kubler2007}
J. K\"{u}bler, G. H. Fecher and C. Felser, Phys. Rev. B
\textbf{76}, 024414 (2007).


\bibitem{Abrikosov}
B. Alling, S. Shallcross, and I. A. Abrikosov, Phys. Rev. B
\textbf{73}, 064418 (2006).


%%%%%%%%%%%%%%%%%%%%%%%%%%%%%%%
\end{thebibliography}
\end{document}